\documentclass[aps,preprint]{revtex4}
\usepackage{graphicx}
\usepackage{epsfig}
\begin{document}

\def \cT {{\cal T}}
\def \cC {{\cal C}}
\def \cD {{\cal D}}
\def \cG {{\cal G}}
\def \cB {{\cal B}}
\def \cU {{\cal U}}
\def \cV {{\cal V}}
\def \cF {{\cal F}}
\def \cT {{\cal T}}
\def \cH {{\cal H}}

\title{Coherent transmission through a one dimensional lattice}

\author{Zhao Yang Zeng$^{1,2}$, Yi-You Nie$^1$, F. Claro$^2$ and W. Yan$^3$}

\affiliation{$^1$Department of Physics, Jiangxi Normal
University, Nanchang 330027, China\\
 $^2$ Facultad de F\'isica, Pontificia Universidad
Cat\'olica de Chile, Casilla 306, Santiago 22, Chile \\
$^3$ Institute of Theoretical Physics, Shanxi University, Taiyuan,
030006, China}

\date{\today}

\begin{abstract}
 Based on the Keldysh nonequilibrium Green function (NGF) technique,
 a general formula for the current and transmission coefficient
 through a one dimensional
 lattice is derived without the consideration of electron-electron interactions.
 We obtain an analytical condition for perfect resonant transmission
 when the levels of sites are aligned, which depends on the parity of the number of
 sites. Localization-delocalization transition in a generic
 one dimensional disordered lattice is also analyzed, depending on
 the correlation among the hopping parameters and the strength of
 the coupling to reservoirs.  The dependence of the number and lineshape of
 resonant transmission and linear conductance peaks on the structure parameters of the
 lattice is also given in several site cases.
\end{abstract}


\maketitle

\section{Introduction}

Transport through a quantum dot lattice depends on the matching of
electron levels in the various quantum dots\cite{gurvitz}. If the
electron levels of individual dots are aligned, resonant tunneling
occurs\cite {wegewijs}. Electronic transport through these
structures is usually investigated theoretically on the basis of a
"classic" rate equation in the weak coupling
limit\cite{zhiming,klimeck}, and a quantum rate equation in the
strong and weak coupling limits\cite{gurvitz,wegewijs}. The
splitting due to the interdot coupling has also been calculated by
Matveev et al.\cite{matveev} and Golden et al.\cite{golden}. Some
groups have recently investigated in detail the spectral and
current properties of a coupled quantum dot pair following the
Keldysh nonequilibrium Green's function (NGF) formalism\cite{niu}.
Calculations on Phonon-assisted\cite{liu} and
photon-assisted\cite{stafford} transport in two coupled quantum
dots have been recently conducted, in which an electron pump
effect is found. Kondo physics and other correlation effects in
the same system have also been considered\cite{aono}. In quantum
dot systems, electron-electron interaction becomes important and
Coulomb oscillation of linear conductance and even Kondo effect
arise.\cite{sohn, Meir}  Experimentally, as early as 1990,
Kouwenhoven et al.\cite{kouwenhoven3} performed a transport
experiment in a sample consisting of fifteen dots. Some
experiments have also been conducted to explore the ground-state
properties of a double quantum dot\cite{wangh}. Very recently,
Oosterkamp et al.\cite{oosterkamp} investigated experimentally
microwave spectroscopy of a two quantum dot molecule.

Recently electronic transport in metallic chains of single atoms
has attracted both theoretical\cite{Liang,Zeng}  and experimental
 \cite{Smit} interest, since atomic wires represent the ultimate
limit of the miniaturization of electrical conductors. Although
atomic wires are somewhat simple toy models in textbooks and
literatures, they may be the best laboratory to test physical
properties of one dimensional ($1D$) systems. The microscopic
mechanism for the conductance quantization of atomic wires still
keeps vague in spite of some endeavours\cite{Yanson}.
First-principle calculations\cite{Liang}  and analytical
deduction\cite{Zeng} demonstrated interesting oscillation of
conduction of the monoatomic wires with the number of atoms, which
was observed experimentally in the conductance of mechanically
controllable break junctions atomic contacts\cite{Smit}. The
parity effect in the conductance was also found in weakly
disordered quasi-1D tight-binding models\cite{Brouwer1} and dirty
superconducting wires\cite{Brouwer2}. However, whether the
interesting parity phenomena arising from single particle
behavior\cite{Zeng} or many-body correlation effects\cite{Oguri}
need more clarifications.

To have a better understanding the transport properties of generic
one diemnsional systems and clarify conductance quantization in
such a system, we consider electron transmission through a one
dimensional lattice where the sites allow for only the
nearest-neighboring hopping. The lattice is linked to reservoirs
through the leads. such a kind of lattice may be one dimensional
quantum dot array or chains of atoms. In this work we adopt the
Keldysh NGF formalism rather than other method such as classic or
quantum rate equation, since the NGF prescription allows one to
obtain explicitly analytical current and transmission formulas,
which are valid in both the strong and weak coupling limits. We
develop a technique to calculate the retarded Green functions for
every site based on the equations-of-motion method. With the help
of the technique we developed, it is very convenient and intuitive
to calculate the retarded and correlation Green functions for
every lattice sites. As the transmission and current formula is
applied to the cases of a single and a double site structure, some
well-known results are rediscovered, with some new phenomena being
observed. The case of triple sites shows more complicated and
interesting resonant transmission structure. If the lattice-lead
couplings are asymmetric, the resonant structure of the
transmission spectra for the triple site structure has weak
dependence on the arrangement of site levels if they are not
aligned. Moreover, we derive an analytical expression for the
condition for perfect transmission of a one dimensional lattice.
Such a condition depends on whether the number of sites is odd or
even, suggesting that the parity feature of the linear conductance
of a one dimensional system does not arise from the
electron-electron correlation effects, since we did not consider
electron-electron interactions. In the case of symmetric
lattice-lead couplings, our result is consistent with the parity
effect of conductance in monatomic wires based on the
first-principles calculations \cite {Liang} and experimental
findings\cite{Smit}.

we notice that Shangguan et al.\cite{Shangguan} studied the
differential conductance and charge distribution in a linear
quantum dot array based on the similar formalism. However, there
exist some crucial discrepancies between our formalisms and
results. First, the prescriptions for the calculation of the
correlation Green functions such as less-than one are different.
After repeating the use of the Keldysh formula in the
noninteraction case with the help of the properties
$G^rG^a=(G^r-G^a)/(2iImG^a)$, we obtain finally the transmission
formula $(23)$, which is different formally from the result of
Shangguan et al.\cite{Shangguan}, Eq. $(27)$ in their paper,
without using the mentioned property. Such a prescription is more
intuitive and facilitates greatly the analysis of the dependence
of transmission probability on the parameters of the system
especially at resonance, as can be seen in the $N$-site case.
Second, we study the linear transport properties such as resonant
transmission, linear conductance,  while Shangguan et al.
investigate the nonlinear ones such as differential conductance,
charge distribution. We also extend the result to conductance
quantization in monovalent atomic chains and the
localization-delocalization transition in a generic one
dimensional system, a fundamental issue in the condensed-matter
physics.

The rest of the paper is organized as follows. In Sec. II we
formulate the transport in a one dimensional lattice and derive
the formulas for calculating the current, transmission and
conductance based on the Keldysh NGF method. In Sec. III we use
the formulas to calculate the transmission and conductance spectra
for a single cite, a double site and a triple site structures,
with a detailed analysis of our results. Also we derive an
analytical expression for the condition for perfect transmission
of $N$-site lattices as the site levels are aligned. Concluding
remarks are given in Sec. IV.

\section{Model and Formulation}

We are interested in the electronic transport properties of a one
dimensional lattice with $N$ sites. The lattice is connected to
the left(right) lead $L$($R$), which can be described by a
tunneling matrix element. The leads are considered electron
reservoirs with a continuum of states filled up to their
respective Fermi levels $\mu_L$ and $\mu_R$ at zero temperature.
Such a device can be described by a single-band tight-binging
model. The on-site energy of site $i$ is labelled by $\epsilon_i$.
An inter-site coupling with matrix element $t_i$ accounts for the
electron's hopping between nearest-neighboring sites $i$ and
$i+1$. The coupling between the left(right) lead $L$($R$) and the
first $1$ (last $N$) site is described by a parameter $v_{kL}$
($v_{kR}$). Then a tight-binding Hamiltonian for the lattice in
the absence of electron-electron interaction takes the following
form
\begin{eqnarray}
H=\sum\limits_{k,\alpha=L,R}\epsilon_{k\alpha}a^+_{k\alpha}a_{k\alpha}+
\sum\limits_{i=1}^{N}\epsilon_id_i^+d_i+\sum\limits_{i=1}^{N-1}(t_id^+_{i+1}d_i+
t^*_id^+_id_{i+1}) \nonumber \\
 +\sum\limits_k(v_{kL}a^+_{kL}d_1+v_{kL}^*d^+_1a_{kL})+
\sum\limits_k(v_{kR}a^+_{kR}d_N+v_{kR}^*d^+_Na_{kR})
\end{eqnarray}
where $a_{kL}$($a_{kL}^+$),$a_{kR}$($a_{kR}^+$) and $d_i$($d^+_i$)
($i=1,2,...N$) are the annihilation (creation) operators for an
electron in the left lead, right leads and at site $i$. In Eq.
$(1)$, the first term is the Hamiltonian for the leads, the second
term is the Hamiltonian for uncoupled $N$ sites, while the third
and last two terms describe the coupling between
neatest-neighboring sites and the couplings between the lattice
and the leads.

Following Ref.\cite{Meir}, the current through the lattice is the
time evolution of the electron number
$N_L(t)=\sum_{k}a_{kL}^+(t)a_{kL}(t)$ in the left lead
\begin{equation}
J_L=-e<\dot N_L>=\frac{ie}{\hbar}<\{N_L,H\}_+>
   =\frac{2e}{\hbar}Re\{\sum\limits_k v_{kL}G_{1,k}^<(t,t)\},
\end{equation}
where the less-than Green function
$G_{1,k}^<(t,t')=i<a^+_{kL}(t')d_1(t)>$. With the help of Dyson's
equation, the less-than Green function $G_{1,k}^<(t,t')$ can be
written as
\begin{equation}
G_{1,k}^<(t,t')=\int
dt_1v_{kL}^*[G_{1,1}^r(tt_1)g^<_{kL}(t_1,t')+G_{11}^<(t,t_1)g^a_{kL}(t_1,t')].
\end{equation}
In Eq. $(3)$, $G_{11}^<(t,t')=i<d^+_1(t')d_1(t)> $, $
G_{11}^r(t,t')=-i\theta(t-t')<\{d_1(t),d_1^+(t')\}_+> $ are the
less-than and retarded Green functions of electrons at site $1$,
while $g_{kL}^{a,<}(t,t')$ is the exact advanced (less-than) Green
function of electrons in the left lead decoupled from the lattice
\begin{eqnarray*}
g_{kL}^{r,a}(t,t')& = & \mp i \theta(\pm t \mp t')e^{-i(\epsilon_{kL}-\mu_L)(t-t')},\\
    g_{kL}^<(t,t')& = & if_L(\epsilon_{kL}-\mu_L)e^{-i(\epsilon_{kL}-\mu_L)(t-t')}.
\end{eqnarray*}
Here $f_{\alpha}(x)=[exp(x/k_BT)+1]^{-1}$ $(\alpha=L, R)$ is the
Fermi-Dirac distribution function of the Lead. Substituting Eq.
$(3)$ into Eq. $(2)$, changing the sum over $k$ into an integral
$\int d\epsilon \rho_L(\epsilon)$ (where $\rho_L(\epsilon)$ the
density of states in the left lead) and introducing a linewidth
function $\Gamma_L(\epsilon)=2\pi \mid v_{kL} \mid^2
\rho_L(\epsilon)$, we get a compact form of current formula in
Fourier space\cite{Meir}
\begin{equation}
J_L=\frac{ie}{\hbar}\int\frac{d\epsilon}{2\pi}\Gamma_L(\epsilon)\{f_L(\epsilon-\mu_L)
[G^r_{11}(\epsilon)-G^a_{11}(\epsilon)]+ G^<_{11}(\epsilon)\}
\end{equation}

Now we turn to calculate the retarded (advanced) Green function
$G^{r,a}_{11}(\epsilon)$ and the less-than Green function
$G^<_{11}(\epsilon)$ of site $1$. The essential idea to solve the
retarded Green function is that we consider first site $2$
neighboring to site $1$ as a part of the right lead, and then
consider site $i+1$ as a part of the right lead when calculating
the Green functions of site $i$. This procedure may be regarded as
an extension of the well-known recursion or decimation method for
calculating Green function.\cite{Ferry} From Dyson's equation
$G=(g^{-1}-\Sigma)^{-1}$, $G^r_{11}(\epsilon)$ can be written as
\begin{equation}
G^r_{11}(\epsilon)=[g^r_{11}(\epsilon)^{-1}-\Sigma^r_{11}(\epsilon)]^{-1},
\end {equation}
where $g^r_{11}(\epsilon)$ is the retarded Green function of site
$1$ decoupled with the system, and
\begin{eqnarray}
\Sigma^r_{11}(\epsilon)&=&\Sigma^r_L(\epsilon)+\Sigma^r_{1R}(\epsilon),\\
\Sigma^r_L(\epsilon)&=&\Sigma_k \mid
v_{kL}\mid^2g^r_{kL}(\epsilon)=\Lambda_L(\epsilon)-\frac{i}{2}\Gamma_L(\epsilon),\\
\Sigma^r_{1R}(\epsilon)&=&\mid t_1\mid^2 G^r_{2R}(\epsilon).
\end{eqnarray}
$G^r_{2R}(\epsilon)$ in Eq. $(8)$ is the retarded Green function
of site $2$ decoupled from site $1$ but still coupled to site $3$.
It is given by
\begin{equation}
G^r_{2R}(\epsilon)=[g^r_{22}(\epsilon)^{-1}-\Sigma^r_{2R}(\epsilon)]^{-1},
\end{equation}
with $\Sigma^r_{2R}(\epsilon)=\mid t_2\mid^2 G^r_{3R}(\epsilon)$.
Similarly, the retarded Green function $G^r_{iR}$ of site $i$ is
\begin{eqnarray}
G^r_{iR}(\epsilon)&=&[g^r_{ii}(\epsilon)^{-1}-\Sigma^r_{i+1R}(\epsilon)]^{-1},\\
\Sigma^r_{iR}(\epsilon)&=&\mid t_i\mid^2
G^r_{i+1R}(\epsilon).\nonumber
\end{eqnarray}
Note that for the last site $N$, its retarded Green function reads
as
\begin{equation}
G^r_{NR}(\epsilon)=[g^r_{NN}(\epsilon)^{-1}-\Sigma^r_R(\epsilon)]^{-1},
\end{equation}
where $\Sigma^r_R(\epsilon)=\Sigma_k \mid
v_{kR}\mid^2g^r_{kR}(\epsilon)=\Lambda_R(\epsilon)-\frac{i}{2}\Gamma_R(\epsilon)$.
Substituting the Green function $G^r_{iR}$ into the exprsession of
the Green function $G^r_{i-1R}$ ($i=2,3,\cdots N$) recursively, we
finally obtain an analytical expression for $G^r_{11}(\epsilon)$
\begin{equation}
G^r_{11}(\epsilon)=\frac{1}{\displaystyle g^r_{11}(\epsilon)^{-1}
                    -\Lambda_L+\frac{i}{2}\Gamma_L
                 -\frac{\mid t_1 \mid^2}{\displaystyle g^r_{22}(\epsilon)^{-1}
                 -\frac{\mid t_2 \mid^2}{\displaystyle \ddots
                 \ddots
                 \frac{\mid t_{N-1}\mid^2} {\displaystyle g^r_{NN}(\epsilon)^{-1}-
                  \Lambda_R+\frac{i}{2}\Gamma_R}}}}.
\end{equation}

Once the retarded and advanced Green functions are known
($G^a=(G^r)^*$), the less-than Green function can be evaluated
with the help of the Keldysh formula for the present problem
\begin{equation}
G^<=G^r\Sigma^<G^a=\frac{G^r-G^a}{1/G^a-1/G^r}\Sigma^<.
\end{equation}
Then
\begin{eqnarray}
G^<_{11}&=&\frac{G^r_{11}-G^a_{11}}{1/G^a_{11}-1/G^r_{11}}\Sigma^<_{11},\\
G^<_{iR}&=&\frac{G^r_{iR}-G^a_{iR}}{1/G^a_{iR}-1/G^r_{iR}}\Sigma^<_{iR},\\
G^<_{NR}&=&\frac{G^r_{NR}-G^a_{NR}}{1/G^a_{NR}-1/G^r_{NR}}\Sigma^<_{R},
\end{eqnarray}
where the self energies are given by
\begin{eqnarray}
\Sigma^<_{11} &=&\Sigma^<_{L}+\Sigma^<_{1R}=\Sigma_k\mid
v_{kL}\mid^2g^<_{kL}(\epsilon)+ \mid t_1\mid^2G^<_{2R}(\epsilon),\\
 &=& if_L( \epsilon-e\mu_L)\Gamma_L+\mid t_1\mid^2G^<_{2R}(\epsilon),\\
\Sigma^<_{iR} &=&\mid t_i\mid^2G^<_{i+1R}(\epsilon), \\
\Sigma^<_{R} &=&\Sigma_k\mid v_{kL}\mid^2g^<_{kL}(\epsilon)=if_R(
\epsilon-\mu_R)\Gamma_R.
\end{eqnarray}
Since $1/G^a-1/G^r=\Sigma^r-\Sigma^a$, combining Eqs. $(14)-(16)$
yields
\begin{equation}
G^<_{11}(\epsilon)=-\frac{f_L(\epsilon-\mu_L)\Gamma_L+f_R(
\epsilon-\mu_R)\Gamma_2}
                   {\Gamma_L+\Gamma_2}[G^r_{11}(\epsilon)-G^a_{11}(\epsilon)],
\end{equation}
where $\Gamma_2=i\mid
t_1\mid^2[G^r_{2R}(\epsilon)-G^a_{2R}(\epsilon)]=-2|t_1|^2ImG^r_{2R}
$. Substituting the above expression  into the current formula
$(4)$, one obtains the following Landauer-B\"uttiker-type
formula\cite{Meir}
\begin{equation}
J_L=-\frac{2e}{\hbar}\int \frac{d\epsilon}{2\pi}[f_L(
\epsilon-\mu_L)-f_R(\epsilon-\mu_R)]
\frac{\Gamma_L\Gamma_2}{\Gamma_L+\Gamma_2}ImG^r_{11}(\epsilon).
\end{equation}
Equation $(22)$ is the central result of this work. This formula
for steady transport is valid both in the strong and  weak
coupling limits. It is also applicable to the nonequilibrium
situation. In Eq. $(22)$, the term
\begin{equation}
\cT(\epsilon)=-\frac{2\Gamma_L\Gamma_2}{\Gamma_L+\Gamma_2}ImG^r_{11}(\epsilon),
\end{equation}
is the transmission coefficient for electron tunneling through the
one-dimensional lattice. Furthermore, the differential conductance
$\cG$ in the linear regime $\mu_L-\mu_R \rightarrow 0^+$ can be
readily derived as
\begin{equation}
\cG =-
 \frac{e^2}{h} \int d \epsilon
\cT(\epsilon)
 \frac{\partial{f(\epsilon-\mu_L)}}{\partial(\epsilon)}
\end{equation}

It is noted that the above procedure to calculate the related
Green functions can be extended to two and three dimension
lattices with each site having multiple levels in the presence of
electron-electron interactions treated in the mean-field
approximation, which will be published elsewhere. In addition,
such a procedure is very suitable to numerical calculation of the
spectral density and tunneling quantities in coupled many site
systems.

\section{Transmission and Conductance}

In this section we will investigate coherent transmission and
conductance in a one dimensional lattice, which may be related to
a one dimensional quantum dot/well array or a chain of single
atoms. Our interest is mainly to find out how parameters determine
the number and position of resonant transmission and conductance
peaks.

\subsection{Single site}

In the case of single site, the Green function and self-energy
take the following simple form
\begin{eqnarray}
G^r_{11}(\epsilon)&=&(\epsilon-\epsilon_1-\Lambda+\frac{i}{2}\Gamma)^{-1},
\nonumber \\
\Gamma_2&=&\Gamma_R,
\end{eqnarray}
where $\Lambda=\Lambda_L+\Lambda_R$, $\Gamma=\Gamma_L+\Gamma_R$.
Then the transmission and conductance are \cite{Meir}
\begin{eqnarray}
\cT (\epsilon)&=&\Gamma_L\Gamma_R \cF_1(\epsilon),
\nonumber \\
\cG  &=& \frac{e^2 \Gamma_L \Gamma_R}{\hbar \Gamma}[
\frac{\partial}{\partial \epsilon_1} Re
f(\epsilon_1+\Lambda-e\mu_L+i\Gamma/2)\nonumber \\
       & & \hspace{1 cm}
+\frac{1}{4{\pi} ^3 k_B T} \sum\limits_{\eta =\pm}
Re\Psi^{(1)}(\frac12+\frac{\eta \Gamma} {4\pi k_B
T}+i\frac{\epsilon_1+\Lambda-e\mu_L}{2\pi k_BT})],
\end{eqnarray}
where
$\cF_1(\epsilon)=[(\epsilon-\epsilon_1-\Lambda)^2+(\Gamma_L+\Gamma_R)^2/4]^{-1}$
, and $\Psi^{(1)}$ is the trigamma function \cite{Abramovitz}. It
is obvious that the transmission coefficient is of the
Breit-Wigner type. Only if the coupling between the site and the
two leads is symmetric, i.e., $\Gamma_L=\Gamma_R$, perfect
resonant transmission ($\cT=1$) can occur at the renormalized
dotsite level $\epsilon_1+\Lambda$, while for asymmetric coupling
$\Gamma_L \ne \Gamma_R$, the transmission coefficient
$\mathcal{T}$ is always less than $1$. The larger the asymmetry,
the smaller  the transmission coefficient. Thus conductance
quantization can be achieved  for the $1$-site system in the
symmetric lattice-lead coupling case.

\subsection{Double sites}

In the case of two coupled sites, it is expected that the
competition between the coupling of the sites and the sites and
leads play a crucial role on the transmission and conductance. In
the absence of coupling to the leads,  the level associated with
two equal sites will split into two levels due to the coupling
between them. The separation between these two split levels is
proportional to the coupling strength. When the sites are
connected to the leads, one should consider the sites and the
leads as a single system, with the result that now the size of the
level splitting depends on both the couplings between sites and
leads.

Consider first a double site structure without coupling to the
leads. Assuming the coupling strength between these two sites is
$t$, the hamiltonian becomes $\epsilon_1 d_1^+d_1+\epsilon_2
d_2^+d_2+td_1^+d_2+t^*d_2^+d_1$. Then the retarded Green functions
for the two sites are
\begin{eqnarray}
G_1^r(\epsilon)&=&
\frac{1}{\epsilon-\epsilon_1-i0^+-|t|^2/(\epsilon-\epsilon_2-i0^+)},\nonumber
\\ G_2^r(\epsilon)&=&
\frac{1}{\epsilon-\epsilon_2-i0^+-|t|^2/(\epsilon-\epsilon_1-i0^+)}.
\end{eqnarray}
One can readily find that the above two retarded Green functions
have the same two poles at
\begin{equation}
\epsilon=\frac{\epsilon_1+\epsilon_2 \pm
\sqrt{(\epsilon_1-\epsilon_2)^2+4|t|^2}}{2}.
\end{equation}
Note that when $\epsilon_1=\epsilon_2$, the level separation is
simply $2|t|$. The above analysis clearly shows that the larger
the coupling, the bigger the separation between the two split
levels. When the sites are connected to the two leads, this new
coupling may modify the effective coupling between the sites as
will be discussed in what follows.

The associated retarded Green functions in the presence of
coupling with the two leads read
\begin{eqnarray}
G^r_{11}(\epsilon) &=&
[\epsilon-\epsilon_1-\Lambda_L+\frac{i}{2}\Gamma_L-|t|^2G^r_{2R}(\epsilon)]^{-1},\\
G^r_{2R}(\epsilon)
&=&(\epsilon-\epsilon_2-\Lambda_R+\frac{i}{2}\Gamma_R)^{-1},
\end{eqnarray}
and
\begin{equation}
 \Gamma_2=\Gamma_R|t|^2/\cB(\epsilon),
\end{equation}
where
$\cB(\epsilon)=(\epsilon-\epsilon_2-\Lambda_R)^2+\Gamma_R^2/4$.
Substituting Eqs. (29)-(31) into the expression $(23)$ yields
\cite{Kawamura}
\begin{equation}
\cT(\epsilon) =\Gamma_L\Gamma_R|t|^2\cF_2(\epsilon)
\end{equation}
where
 \begin{equation} \cF_2(\epsilon)=\frac{\cB(\epsilon)}{[
 (\epsilon-\epsilon_1-\Lambda_L)\cB(\epsilon)-|t|^2
(\epsilon-\epsilon_2-\Lambda_R)]^2+[\Gamma_L\cB(\epsilon)
+|t|^2\Gamma_R]^2/4},
\end{equation}

 We now
consider the case when the levels of the isolated sites $1$ and
$2$ are the same ($\epsilon_1=\epsilon_2=\epsilon_0$). In the case
of symmetric coupling, i.e., $\Gamma_L=\Gamma_R=\Gamma$,
$\Lambda_L= \Lambda_R=\Lambda$, the condition for perfect resonant
transmission is that
\begin{equation}
(\epsilon-\epsilon_0-\Lambda)^2+\Gamma^2/4=|t|^2
\end{equation}
has real roots. This equation also determines the number and
position of the resonant transmission peaks. Obviously, the
condition for perfect resonant transmission is $|t|^2 \ge
\Gamma^2/4$. There will be then only one perfect resonant
transmission peak located at $\epsilon=\epsilon+\Lambda$ when
$|t|^2=\Gamma^2/4$. If $|t|^2 < \Gamma^2/4$ , there is just one
imperfect vresonant peak ($\cT<1$) pinned at
$\epsilon=\epsilon_0+\Lambda$. In the case of $|t|^2
> \Gamma^2/4$, two perfect transmission peaks exist at
$\epsilon_{\pm}=\epsilon_0+\Lambda \pm \sqrt{|t|^2-\Gamma^2/4}$.
These features can be clearly seen in  Fig. $1$ (a), where we can
also appreciate that the lineshape of all the transmission peaks
is Lorenzian. Here and in all figures following energies are in
arbitrary units.

Next we consider what would happen if the coupling between the
sites and the two leads becomes asymmetric. In this case, one can
find from the expression of the transmission coefficient $(32)$
that the condition for perfect transmission is $|t|^2=\Gamma_L
\Gamma_R/4$ and only one perfect transmission peak can be expected
at $\epsilon=\epsilon_0+\Lambda$. In all other cases, the
transmission coefficient is less than $1$. In Fig. $1$ (b) we show
the transmission coefficient for various inter-site coupling
constants $t$ for $\Gamma_L=0.25$ and $\Gamma_R=4$. If the level
shift $\Lambda_L$ and $\Lambda_R$ induced by the left and right
lead are not the same, the lineshape of the transmission peaks is
non Lorenzian, as is apparent in the case $\Lambda_L=0.025$,
$\Lambda_R=0.4$ also shown in the figure.
 Figure $1$ (c) and
(d) are the plots of the transmission coefficient for increasing
inter-site coupling $t$ in the symmetric and asymmetric coupling
cases, respectively. When the levels of the sites are not aligned.
It is then expected that no perfect resonant transmission exists.
One can observe that two symmetric transmission peaks always
resolved in the symmetric coupling case. As the inter-site
coupling $t$ increases, the maximum value as well as the
separation of the transmission peaks increases. For asymmetric
coupling, one transmission peak can be expected when the coupling
between the sites is weak, and two asymmetric transmission peaks
when it is strong. In addition, with increasing interdot coupling
$t$, the value of the transmission coefficient increases first and
then decreases after it is saturated. The position of the
principal transmission peak is closer to the level of the site
with smaller coupling to the lead.

This complex behavior reflects the competition between two
resonances derived from the ground state of each site when far
apart. As they come together and when coupling increases, the
resonances approach the ground and first excited states of the
compound system, which in general separated in energy. If the
coupling is weak, on the other hand, the peaks will only be
resolved if their width is smaller than the energy difference
$\Delta \epsilon=|\epsilon_2-\epsilon_1|$. If the couplings to the
leads is asymmetric the competing resonances will be dominated by
that whose associated wavefunction is concentrated on the site
region must weakly coupled to the lead.

To sum  up, we have investigated in detail the resonant structure
of the transmission coefficient of a coupled double site structure
in the linear regime. We found that when the levels of the two
sites are aligned, the condition for perfect resonant transmission
is $|t|^2 \ge \Gamma^2/4$ in the case of symmetric site-lead
coupling, and $|t|^2 =\Gamma_L \Gamma_R/4$ in the asymmetric case,
which is consistent with the derivations of Ref. \cite{Larkin}.
Once there exists a mismatch between the two levels of the sites,
no perfect resonant transmission peak can be expected. In the case
of symmetric site-lead coupling, the value of the transmission
coefficient increases with increasing inter-site coupling, until
it is saturated. For asymmetric site-lead coupling, the
transmission coefficient increases first and then decreases after
it reaches a maximum, as the inter-site coupling is increased.
Moreover, in the asymmetric case, the splitting of the
transmission peak requires a stronger inter-site coupling $t$.
Asymmetry between the two split transmission peaks can be observed
when the two levels are not aligned for asymmetric site-lead
coupling, provided the inter-site coupling is strong enough. In
addition, different level shifts of the two sites induced by the
coupling to the leads will introduce a non Lorenzian lineshape of
the transmission peak(s). The results imply that \emph{no perfect
transmission and then no conductance quantization can be expected
for double site systems in the symmetric coupling case(inversion
symmetry)}.

\subsection{Triple Sites}

A coupled triple site structure is expected to contain richer
physics than a coupled double site system,  since it permits more
interesting arrangement of the energy levels in each site and the
competition between inter-site couplings. Surprsingly, electronic
transport through the triple sites has been less investigated in
the past both theoretically and experimentally\cite{Lee}. In this
subsection, we study the detailed dependence of the number and
profile of transmission peaks, on the parameters of such a
structure.

For an isolated triple site system, the hamiltonian can be written
as
\begin{equation}
H=\epsilon_1 d_1^+d_1+\epsilon_2 d_2^+d_2+\epsilon_3
d_3^+d_3+t_Ld^+_1d_2+t^*_Ld_2^+d_1+t_Rd^+_2d_3+t^*_Rd_3^+d_2,
\end{equation}
where $t_L$ and $t_R$ are the couplings between sites $1$ and $2$,
and $2$ and $3$, respectively. The related retarded Green
functions are readily derived as
\begin{eqnarray}
G_1^r(\epsilon)&=&\frac{1}{\epsilon-\epsilon_1-i0^+
-\frac{|t_L|^2}{\epsilon-\epsilon_2-i0^+-\frac{|t_R|^2}{
\epsilon-\epsilon_3-i0^+}}}, \nonumber \\
G_2^r(\epsilon)&=&
\frac{1}{\epsilon-\epsilon_2-i0^+-|t_L|^2/(\epsilon-\epsilon_1-i0^+)
-|t_R|^2/(\epsilon-\epsilon_3-i0^+)}, \nonumber \\
G_3^r(\epsilon)&=&\frac{1}{\epsilon-\epsilon_3-i0^+
-\frac{|t_R|^2}{\epsilon-\epsilon_2-i0^+-\frac{|t_L|^2}{
\epsilon-\epsilon_1-i0^+}}} .
\end{eqnarray}
when the levels of three sites are aligned, i.e.,
$\epsilon_1=\epsilon_2=\epsilon_3=\epsilon_0$, each of these three
retarded Green functions has the same three poles at
$\epsilon=\epsilon_0$,
$\epsilon=\epsilon_0+\sqrt{|t_L|^2+|t_R|^2}$ and
$\epsilon=\epsilon_0-\sqrt{|t_L|^2+|t_R|^2}$.  When the three
electron levels are mismatched, there exist in general three poles
in the above retarded Green functions, that is to say, the
position of the original levels are moved due to the existence of
the inter-site  couplings. As the couplings between the structure
and the leads are turned on, the Green functions for calculating
the transmission coefficient become
\begin{eqnarray}
G^r_{11}(\epsilon) &=&
[\epsilon-\epsilon_1-\Lambda_L+\frac{i}{2}\Gamma_L-|t_L|^2G^r_{2R}(\epsilon)]^{-1},\\
G^r_{2R}(\epsilon)
&=&(\epsilon-\epsilon_2-|t_R|^2G^r_{3R})^{-1},\\
G^r_{3R}(\epsilon)
&=&(\epsilon-\epsilon_3-\Lambda_R+\frac{i}{2}\Gamma_R)^{-1},
\end{eqnarray}
and
\begin{equation}
\Gamma_2=-2|t_L|^2ImG^r_{2R}.
\end{equation}
Then the transmission coefficient becomes
\begin{equation}
\cT(\epsilon) =\Gamma_L\Gamma_R|t_L|^2|t_R|^2\cF_3(\epsilon),
\end{equation}
where
\begin{equation}
\cF_3(\epsilon)=\frac{\cB(\epsilon)}{\cC(\epsilon)
[(\epsilon-\epsilon_1-\Lambda_L-|t_L|^2ReG^r_{2R})^2+(\Gamma_L-2|t_L|^2
ImG^r_{2R})^2/4]}
\end{equation}
with
\begin{eqnarray}
\cB(\epsilon)&=&(\epsilon-\epsilon_3-\Lambda_R)^2+\Gamma_R^2/4,\nonumber \\
\cC(\epsilon)&=&[(\epsilon-\epsilon_2)\cB(\epsilon)-|t_R|^2
(\epsilon-\epsilon_3-\Lambda_R)]^2+|t_R|^4\Gamma_R^2/4, \nonumber
\end{eqnarray}
and
\begin{eqnarray}
Re G^r_{2R}(\epsilon)&=&[(\epsilon-\epsilon_2)\cB
(\epsilon)-|t_R|^2(\epsilon-\epsilon_3-\Lambda_R)]\frac{\cB(\epsilon)}{\cC(\epsilon)}, \nonumber \\
Im G^r_{2R}(\epsilon)&=&-\frac{|t_R|^2\Gamma_R}{2}
\frac{\cB(\epsilon)}{\cC(\epsilon)}.
\end{eqnarray}
Now we consider the case when the three levels of sites are
aligned and neglect the energy shifts $\Lambda_L$ and $\Lambda_R$.
At resonance $\epsilon=\epsilon_0$, one has
\begin{eqnarray}
\cB(\epsilon)&=&\Gamma_R^2/4,\nonumber \\
\cC(\epsilon)&=&|t_R|^4\Gamma_R^2/4, \nonumber \\
Im G^r_{2R}(\epsilon)&=&-\frac{\Gamma_R}{2|t_R|^2},
\end{eqnarray}
then one can easily derive the condition for perfect transmission
\begin{equation}
\mid\frac{t_L}{t_R}\mid^2=\frac{\Gamma_L}{\Gamma_R}.
\end{equation}
Further analytical results in this case are cumbersome, so we
shall next provide some numerical results on the transmission
spectra of the coupled triple site structure in various cases.
Figures $2$ and $3$ show the transmission spectra for symmetric
coupling ($\Gamma_L=\Gamma_R=1$), and asymmetric coupling
($\Gamma_L=1, \Gamma_R=4$)  to the left and right leads,
respectively. In the figures four kinds of arrangements of energy
levels  are presented: (a) aligned levels
$\epsilon_1=\epsilon_2=\epsilon_3=10$, (b) ladder levels
$\epsilon_1=9$, $\epsilon_2=10$, $\epsilon_3=11$, (c) $V$-type
levels $\epsilon_1=9$, $\epsilon_2=11$, $\epsilon_3=9$ and
 (d) two aligned neighboring levels
$\epsilon_1=9$, $\epsilon_2=9$, $\epsilon_3=11$. From Fig. 2 (a),
one finds that there exists only one perfect transmission peak
located at $\epsilon=\epsilon_0$ when the inter-site couplings are
equal and small. As $t_L=t_R$ are increased, the perfect
transmission peak splits symmetrically. Once the energy shifts are
taken into consideration, the transmission peaks are no longer
equally spaced (dotted line in Fig. 2 (a)). Since Eq. $(45)$ is no
longer satisfied, there is no perfect transmission when $t_L \ne
t_R$.

In the case of levels arranged in a ladder sequence (Fig. 2 (b)),
one perfect transmission peak with two low shoulders is obtained
for $t_L=t_R=0.5$. As they both reach the value $2$, the shoulders
become two imperfect transmission peaks and the perfect
transmission peak at the center is widened. When the inter-site
couplings are not equal, the transmission peaks are suppressed and
arranged into a ladder type.  When the three site levels are in
the $V$-type disposition, interesting transmission spectra are
observed. If the inter-site couplings are equal and small, one
sharp perfect transmission peak and one broad imperfect
transmission peak are seen. As the inter-site couplings increase,
the imperfect transmission peak splits and the resonances move
further apart.
 Similarly,
the inequality of $t_L$ and $t_R$ decreases the transmission in
all three peaks. In fact the value of the transmission coefficient
through split electron levels is determined by the extension to
the leads of the wavefunctions for these split electron levels.
Then the above phenomena can be similarly explained as we did in
the case of two coupled sites. Figure 2 (d) shows the transmission
spectra for $\epsilon_1=9$, $\epsilon_2=9$, $\epsilon_3=11$. One
can see two asymmetric imperfect transmission peaks as
$t_L=t_R=0.5$. When they reach the value $2$, three imperfect
transmission peaks of different height are discerned. As in the
case of three ladder levels, the asymmetry between two inter-site
couplings $t_L$ and $t_R$ rearranges these three transmission
peaks into a ladder. Notice that no perfect transmission peak even
appears in this case.

Now we investigate how the transmission spectra  are modified when
the site-lead couplings are asymmetric. One can find from Fig. 3
that perfect transmission exists only in the case of three aligned
 levels under the condition for perfect transmission. Figure
3 (a) is the transmission spectra for different inter-site
couplings when the levels of three sites are aligned. One
imperfect transmission peak is seen as $t_L=t_R$ and one perfect
transmission peak along with two narrow shoulders is observed as
$t_L=1$, $t_R=2$, which satisfies the condition for perfect
transmission $(45)$. Also one can find different resonant
structures for the cases $|t_L/t_R|^2
> \Gamma_L / \Gamma_R$ and $|t_L/t_R|^2 \le \Gamma_L / \Gamma_R$.
When the three levels are not aligned, no perfect transmission
occurs in the case of asymmetric site-lead couplings, which is
demonstrated in Fig. 3 (b)-(d). The transmission spectra in the
case of levels arranged in a ladder exhibit one imperfect
transmission peak and one shoulder structure as $t_L=t_R=0.5$.
Increasing the inter-site couplings, the shoulder becomes
gradually distinguished as a transmission peak. The competition
between $t_L$ and $t_R$ strongly modifies the resonant structure
of the transmission spectra. Comparing  Fig. 3 (b)-(d), one may
find that the resonant structure of the transmission spectra are
similar and mainly display two transmission peaks structure. It
suggests a fact that, as the site-lead couplings are not
symmetric, the resonant structure of the transmission spectra has
weak dependence on the arrangement of the electron levels of sites
if they are not aligned.

From the above analysis, the transmission spectra of three coupled
sites is more complicated and contains more physics than that of
two coupled sites. The main features of the transmission spectra
for a triple site structure are as follows: (1) the transmission
spectra may contain just one, two or three resonant peaks, (2) the
resonant structure of the transmission spectra is strongly
dependent on the arrangement of the electronic levels when the
site-lead couplings are symmetric, while weak dependence on the
arrangement of the electron levels if they are not, (3) the
resonant structure of the transmission spectrum strongly depends
on the symmetry between inter-site couplings, (4) \emph{perfect
transmission and thus conductance quantization can be achieved for
a triple site system with inversion symmetry when coupled
symmetrically to reservoirs.} These conclusions hold for the
linear conductance $\cG$ as well, since the conductance and the
transmission coefficient possess the same resonant structure at
low temperatures. Our theoretical results are consistent with the
experiment observations in the conductance of triple
sites\cite{Lee}.

\subsection{N Sites}

As can be found from our previous analysis, the condition for
perfect transmission is different for one, two and three site
structures. It gives us a hint that \emph{the condition for
perfect transmission would depend on whether the number of sites
in a 1D lattice is even or odd}. In what follows, we derive the
condition under which the perfect transmission happens when the
electron levels of the lattice are aligned and energy shifts are
ignored, i.e.,
$\epsilon_1=\epsilon_2=\cdots=\epsilon_N=\epsilon_0$, and
$\Lambda_L=\Lambda_R=0$.  From the expression for the transmission
coefficient (23) and for $\Gamma_2$, one has
\begin{eqnarray}
\cT(\epsilon)&=&\frac{4\Gamma_L|t_1|^2ImG^r_{2R}}{\Gamma_L-2|t_1|^2ImG^r_{2R}}
              ImG^r_{11}(\epsilon) \nonumber \\
          &=& \frac{-2\Gamma_L|t_1|^2ImG^r_{2R}}
              {(\epsilon-\epsilon_0-|t_1|^2ReG^r_{2R})^2+
              (\Gamma_L-2|t_1|^2ImG^r_{2R})^2/4}.
\end{eqnarray}
At resonance $\epsilon=\epsilon_0$, the real part of all the
retarded Green functions becomes zero. Then the transmission
coefficient is given by
\begin{equation}
\cT=\frac{-8\Gamma_L|t_1|^2ImG^r_{2R}}
              {(\Gamma_L-2|t_1|^2ImG^r_{2R})^2} .
\end{equation}
Perfect transmission through the lattice, $\cT=1$, is obtained if
\begin{equation}
\Gamma_L=-2|t_1|^2ImG^r_{2R}.
\end{equation}
Notice that since $g^r_{ii}=0$ $(i=2,3,\cdots N)$ at resonance
$\epsilon=\epsilon_0$, one can obtain the following expression for
$ImG^r_{2R}$
\begin{eqnarray}
ImG^r_{2R}&=&-\frac{\Gamma_R}{2}|\frac{t_3t_5\cdots
                t_{N-2}}{t_2t_4\cdots t_{N-1}}|^2; \hspace{0.2cm}N\hspace{0.1cm}
                 odd\\
ImG^r_{2R}&=&-\frac{2}{\Gamma_R}|\frac{t_3t_5\cdots
                t_{N-1}}{t_2t_4\cdots t_{N-2}}|^2; \hspace{0.2cm}N\hspace{0.1cm}
                 even.
\end{eqnarray}
Then the condition for perfect transmission is
\begin{eqnarray}
|\frac{t_1t_3\cdots t_{N-2}}{t_2t_4\cdots t_{N-1}}|^2 &=&
\frac{\Gamma_L}{\Gamma_R}; \hspace{0.8cm}N\hspace{0.1cm}
                 odd\\
|\frac{t_1t_3\cdots t_{N-1}}{t_2t_4\cdots t_{N-2}}|^2 &=&
                  \frac{\Gamma_L \Gamma_R}{4}; \hspace{0.3cm}N\hspace{0.1cm}
                 even.
\end{eqnarray}
If the interdot couplings are the same, the condition becomes
\begin{eqnarray}
 \Gamma_L&=&\Gamma_R; \hspace{0.8cm}N\hspace{0.1cm}odd\\
 |t_{\frac{N}{2}}|^2 &=&\frac{\Gamma_L \Gamma_R}{4}; \hspace{0.3cm}N\hspace{0.1cm}
                 even.
\end{eqnarray}

Thus one can conclude that for a one dimensional  lattice, the
condition for perfect transmission is dependent on the parity of
the sites of the lattice, i.e., whether the number of sites is odd
or even. Equation (51) suggests that for a chain with an odd
number of sites and inversion symmetry (i.e., $\Gamma ^{L}=\Gamma
^{R}=\Gamma, t_1=t_{N-1},t_2=t_{N-2},$ etc.) perfect transmission
will be  automatically satisfied at the middle of the band or
level group. Thus the linear conductance is quantized to the value
$2e^{2}/h. $ This is not the case when $N$ is even.  From the
 Eqs. (52) and (46), we find a transmission
coefficient $4\eta/(1+\eta)^2$ with $\eta=|\frac{2t_1t_3\cdots
t_{N-1}}{\Gamma t_2t_4\cdots t_{N-2}}|^2$ less than unity and a
conductance smaller than $2e^{2}/h$.  The even-odd feature appears
in transmission and conductance in the absence of any electron
interactions, thus \emph{the parity feature is not due to many
particle effects}.  Our argument also proves that when the system
is symmetric under inversion, the state at the middle of the band
is always delocalized, regardless of the amount of disorder that
respects such symmetry condition. Eqs. (51) and (52) defines a
broad class of correlations in the disorder yielding
delocalization of $1D$ disordered systems.  This parity effect of
conductance is consistent with the predictions in the monatomic
wires based on the first-principles calculation\cite{Liang} and
confirmed by the experimental observations\cite{Smit}.

    The dependence of transmission or conductance on the number of
    sites can be explained qualitatively as follows. To
    simplify our discussion, we constrain ourselves to the case when
    the inter-site couplings are the same, which corresponds to
    the monovalent atomic wire case\cite{Liang}. When the number $N$  of sites is odd,
    one finds that the current through the middle site $(N+1)/2$
     can be written as
     \begin{equation}
     J_o=-\frac{2e}{\hbar}\int \frac{d\epsilon}{2\pi}[f_L(\epsilon-\mu'_L)-f_R(\epsilon-\mu'_R)]
     \frac{\Gamma_{LD}\Gamma_{RD}}{\Gamma_{LD}+\Gamma_{RD}}
     ImG^r_{\frac{N-1}{2}\frac{N-1}{2}}(\epsilon),
     \end{equation}
  where $\Gamma_{LD}=-2t^2ImG^r_{\frac{N-1}{2}L}$,
 $\Gamma_{RD}=-2t^2ImG^r_{\frac{N+3}{2}R}$, with
 $G_{\frac{N-1}{2}L}$ and $G_{\frac{N+3}{2}R}$ are the Green's
 functions decoupled from the middle site $(N+1)/2$. In Eq. (55) we
 denote the chemical potential of the site left(right) to the
 middle site by $\mu'_L$ ($\mu'_R$) which equals to the chemical potential
 $\mu_L$ ($\mu_R$) in the equilibrium. The conductance at resonance can then be
 expressed as
    \begin{equation}
     \cG_o=\frac{2e^2}{h}
     \frac{4\Gamma_{LD}\Gamma_{RD}}{(\Gamma_{LD}+\Gamma_{RD})^2}.
     \end{equation}
     Obviously, the conductance reaches $\frac{2e^2}{h}$ only if
$\Gamma_{LD}=\Gamma_{RD}$, which implies $\Gamma_L=\Gamma_R$ from
the expressions of $\Gamma_{LD}$ and $\Gamma_{RD}$. When the site
number $N$ is even, one can divide the lattice into two parts and
then the current from the site $N/2$ to $(N+2)/2$ can be written
as \cite{Cuevas}
\begin{equation}
  J_e=\frac{2e}{\hbar}\int \frac{d\epsilon}{2\pi}[f_L(\epsilon-\mu'_L)-f_R(\epsilon-\mu'_R)]
   \frac{4t^2ImG^r_{\frac{N}{2}L}ImG^r_{\frac{N+2}{2}R}}
  {|1-t^2G^r_{\frac{N}{2}L}G^r_{\frac{N+2}{2}R}|^2}.
\end{equation}
The conductance at resonance ($ReG^r=0$) is
 \begin{equation}
     \cG_e=\frac{2e^2}{h}\frac{4t^2ImG^r_{\frac{N}{2}L}ImG^r_{\frac{N+2}{2}R}}
  {|1+t^2ImG^r_{\frac{N}{2}L}ImG^r_{\frac{N+2}{2}R}|^2}.
 \end{equation}
 If $ \cG_e=\frac{2e^2}{h}$, one needs
 $t^2=(ImG^r_{\frac{N}{2}L}ImG^r_{\frac{N+2}{2}R})^{-1}$, which eventually
 yields $t^2=\Gamma_L\Gamma_R/4$ after simple calculation. It is
 in fact the perfect transmission condition  in the case of double
 sites. Physically the even-odd dependence of conductance
 reflects the different conditions of constructive interference in
 the different circumstances. For the lattice with  odd number of sites, one can view the array as a single site
  coupled with two
 renormalized leads. If the number of sites is even, the lattice
 can be considered as a two renormalized normal metal contact or a
 double sites coupled with two renormalized leads.
 The condition for perfect transmission is more strict in
 the later case.

 Finally, we would like discuss  possible effects arising from electron-electron
 interaction. It is expected that the main results may hold as well.
 In the presence of on-site Coulomb interaction, an additional term
 of self-energy will be introduced for the Green function of each
 site.  The effect is just shifting the site level and splitting the resonance
 position at absolute zero temperature, since elastic scattering
 can only be expected at zero temperature.\cite{Langreth} The
 central formula $(23)$ remains formally the same in the presence
 of interactions as long as the ground-state of the system possess Fermi liquid properties.\cite{Rejec}

\section{Conclusions}

In summary, using the Keldysh nonequilibrium Green function
method, we have derived the formulas to calculate the transmission
coefficient, current and conductance of a chian of coupled $N$
site system. An effective and convenient procedure to calculate
recursively the retarded(advanced) and lesser(greater) Green
functions has been developed. Based on the formulation developed,
we have analyzed the transmission spectra of just single site,
double site and triple site structures in detail, obtaining some
well-know results and finding some new features in the
transmission spectra of  double and triple site systems. When the
electron levels of $N$ sites are aligned, we have obtained an
analytical expression for the condition for perfect transmission,
demonstrating the even-odd parity effect in the transmission of a
generic one dimensional lattices with inversion symmetry.

\begin{center}
{\bf ACKNOWLEDGMENT}
\end{center}
This work was supported in part by a C\'atedra Presidencial en
Ciencias, FONDECYT 1020829 of Chile and excellent talent fund of
Jiangxi Normal University.


\begin{figure}
\caption{Transmission spectra of a double site structure in the
case of symmetric (a,c) and asymmetric (b,d) site-lead couplings.
}
\end{figure}

\begin{figure}
\caption{Transmission spectra of a  triple site structure in the
case of symmetric site-lead couplings ($\Gamma_L=\Gamma_R=1$):
(a)$\epsilon_1=\epsilon_2=\epsilon_3=10$; (b)$\epsilon_1=9,
\epsilon_2=10, \epsilon_3=11$; (c)$\epsilon_1=9, \epsilon_2=11,
\epsilon_3=9$ and (d)$\epsilon_1=9, \epsilon_2=9, \epsilon_3=11$.}
\end{figure}

\begin{figure}
\caption{Transmission spectra of a triple site structure in the
case of asymmetric site-lead couplings ($\Gamma_L=1,\Gamma_R=4$):
(a)$\epsilon_1=\epsilon_2=\epsilon_3=10$; (b)$\epsilon_1=9,
\epsilon_2=10, \epsilon_3=11$; (c)$\epsilon_1=9, \epsilon_2=11,
\epsilon_3=9$ and (d)$\epsilon_1=9, \epsilon_2=9, \epsilon_3=11$.}
\end{figure}


\begin {references}

\bibitem {gurvitz} S. A. Gurvitz, Phys. Rev. B {\bf 57}, 6602
(1998).
\bibitem {wegewijs} M. R. Wegewijs and Yu. V. Nazarov, Phys. Rev. B
{\bf 60}, 14318 (1999).
\bibitem {zhiming} Z. M. Yu et al., Phys. Rev. B {\bf 55},13697
(1997); {\bf 58}, 13830 (1998).
\bibitem {klimeck} G. Klimeck et al., Phys. Rev. B {\bf 50}, 2316
(1994); G. Chen et al., ibid. {\bf 50}, 8035 (1994);
\bibitem {matveev} K. A. Matveev, Phys. Rev. B {\bf 51}, 1743
(1995); K. A. Matveev et al., ibid. {\bf 53}, 1034 (1996).
\bibitem {golden} J. M. golden et al., Phys. Rev. B {\bf 53}, 3893
(1996); {\bf 54}, 16757 (1996).
\bibitem {niu} C. Niu et al., Phys. Rev. B {\bf 51}, 5130 (1995);
J. Q. You et al., ibid. {\bf 60}, 8727, 13314 (1999); S. Lamba and
S. K. Joshi, ibid. {\bf 62}, 1580 (2000).
\bibitem {liu} L. J. Liu et al., Phys. Rev. B {\bf 54}, 1953 (1996).
\bibitem {stafford} C. A. Stafford and N. S. Wingreen, Phys. Rev. Lett. {\bf 76}, 1916 (1996);
T. H. Stoof and Yu. V. Nazarov, Phys. Rev. B {\bf 53}, 1050
(1996); Q. F. Sun et al., ibid. {\bf 61}, 12643 (2000); Z. S. Ma
et al., Phys. Rev. B {\bf 62}, 15352 (2000).
\bibitem {aono} T. Aono et al., J. Phys. Soc. Jpn. {\bf 67}, 1860
(1998); A. Georges and Y. Meir, Phys. Rev. Lett. {\bf 82}, 3508
(1999).
\bibitem {sohn} Lydia L. Sohn, Leo P. Kouwenhoven, Gerd Schon,
 {\it Mesoscopic Electron Transport} (Nato Asi Series. Series E, Applied Sciences,
 No. 345, Kluwer Academic Publishers,1997).

\bibitem {Meir} Y. Meir and N.S. Wingreen, Phys. Rev. Lett. {\bf 68}, 2512 (1992);
S. Hershfield, J.H. Davies, and J.W. Wilkins, ibid. 67, 3720
(1991); A.P. Jauho, N.S. Wingreen, and Y. Meir, Phys. Rev. B 50,
5528 (1994).
\bibitem {kouwenhoven3} L. P. Kouwenhoven et al., Phys. Rev. Lett.
{\bf 65}, 361 (1990).
\bibitem {wangh}  F. Hofmann et al., Phys. Rev. B {\bf 51}, 13872 (1995); R.
Blick et al., ibid. {\bf 53}, 7899 (1996); D. Dixon et al., ibid.
{\bf 53}, 12625 (1996). T. Schmidt et al., Phys. Rev. Lett. {\bf
78}, 1544 (1997).
\bibitem {oosterkamp} T. H. Oosterkamp et al., Nature {\bf 395},
873 (1998).
\bibitem{Liang} N. D. Liang, Phys. Rev. Lett. {\bf 79} 1357
(1997); N. D. Lang and Ph. Avouris, Phys. Rev. Lett. {\bf 81},
3515 (1998); H. -S. Sim, H. -W. Lee, and K. J. Chang, Phys. Rev.
Lett. {\bf 87}, 096803 (2001); S. Tsukamoto and K. Hirose, Phys.
Rev. B {\bf 66}, 161402 (2002).
\bibitem{Zeng} Z. Y. Zeng and F. Claro, Phys. Rev. B {\bf 65},
193405 (2002).
\bibitem {Smit} R. H. M. Smit, C. Untiedt, G. Rubio-Bollinger, R.C. Segers, and
J.M. van Ruitenbeek, Phys. Rev. Lett. {\bf 91}, 076805 (2003); K.
S. Thygesen and K.W. Jacobsen, hys. Rev. Lett. {\bf 91}, 146801
(2003).
\bibitem {Yanson} A. I. Yanson, G. Rubio-Bollinger, H. E. van den Brom,
N. Agrat, and J.M. van Ruitenbeek, Nature (London) 395, 783
(1998); A.I. Yanson and J.M. van Ruitenbeek, Phys. Rev. Lett. {\bf
79} 2157 (1997); J.M. van Ruitenbeek, M.H. Devoret, D. Esteve and
C. Urbina, Phys. Rev. B {\bf 56}  12566 (1997); J. C. Cuevas, A.
Levy Yeyati and A. Martin-Rodero Phys. Rev. Lett. {\bf 80}, 1066
(1998).
\bibitem{Brouwer1} P. W. Brouwer, C. Mudry, B. D. Simons,
and A. Altland, Phys. Rev. Lett. {\bf 81}, 862 {1998}.
\bibitem{Brouwer2}P. W. Brouwer, A. Furusaki, I. A. Gruzberg, and C. Mudry,
Phys. Rev. Lett. {\bf 85}, 1064 (2000).
\bibitem {Oguri} A. Oguri, Phys. Rev. B {\bf 59}, 12240 (1999); {\bf 63}, 115305 (2001).
\bibitem {Shangguan} W. Z. Shangguan, T. C. Au Yeung, Y. B. Yu, and C. H.
Kam, Phys. Rev. B {\bf 63}, 235323 (2001).
\bibitem {Ferry} D. Ferry and S. M. Goodnick, {\it Transport in
Nanostructures}, Cambridge University Press, 1999.
\bibitem {Abramovitz} M. Abramovitz and I. A. Stegun, {\it Handbook of
Mathematical Functions}, Dover, New York (1972).
\bibitem {Kawamura} K. Kawamura and T. Aono, Jpn. J. Appl. Phys.
Part I {\bf 36}, 3951 (1997).
\bibitem {Larkin} A. I. Larkin and K. A. Matveev, Zh. Eksp. Teor.
Fix. {\bf 93}, 1030 (1987) [Sov. Phys. JETP {\bf 66}, 580 (1987)];
P. Pals and A. MacKinnon, J. Phys. Condens. Matter {\bf 8}, 5401
(1996).
\bibitem {Lee} S. D. Lee et al., Phys. Rev. B {\bf 62}, R7735
(2000); F. Waugh et al., Phys. Rev. Lett. {\bf 75}, 705 (1995); R.
Kotlyar and S. Das Sarma, Phys. Stat. Sol. (b) {\bf 204}, 335
(1997); M. R. Wegewijs, Yu. V. Nazarov and S. A. Gurvitz, {\it
Interaction effects in semiconductor one-dimensional systems}, Ed.
by T. Brandes, Lecture Notes in Physics (Springer, 2000).
\bibitem {Cuevas} J. C. Cuevas, A. Martin-Redero, and A. Levy
Yeyati, Phys. Rev. B {\bf 54}, 7366 (1996).
\bibitem{Langreth} D. C. Langreth, Phys. Rev. {\bf 150}, 516 (1966).
\bibitem{Rejec} T. Rejec and A. Ramsak, Phys. Rev. B {\bf 68}, 035342 (2003).








\end{references}
\end{document}